\documentclass[twocolumn,prl,showpacs,preprintnumbers,amsmath,amssymb,floatfix]{revtex4}
\usepackage{graphicx}
\usepackage{amsmath}
\usepackage{amssymb}
\usepackage{color}

\begin{document}

\title{Optimal Percolation of Disordered Segregated Composites}

\author{Niklaus Johner}\email[Email: ]{niklaus.johner@epfl.ch}
\author{Claudio Grimaldi}\email[Email: ]{claudio.grimaldi@epfl.ch}
\author{Thomas Maeder}
\author{Peter Ryser}

\affiliation{LPM, Ecole Polytechnique F\'ed\'erale de
Lausanne, Station 17, CH-1015 Lausanne, Switzerland}

%\date{15-09-04}

%\widetext

\begin{abstract}
We evaluate the percolation threshold values for a realistic model of continuum segregated systems,
where random spherical inclusions forbid the percolating objects, modellized by hard-core spherical
particles surrounded by penetrable shells, to occupy large regions inside the composite.
We find that the percolation threshold is generally a non-monotonous function of segregation,
and that an optimal (i. e., minimum) critical concentration exists well before maximum segregation is reached.
We interpret this feature as originating from a competition between reduced available volume effects and
enhanced concentrations needed to ensure percolation in the highly segregated regime.
The relevance with existing segregated materials is discussed.
\end{abstract}
\maketitle

The percolation threshold of a two-phase heterogeneous system denotes the
critical concentration at which global (long-range) connectivity of one phase is
first established, and is accompanied by a sudden transition of the effective properties of
the whole system, such as the conductivity in conductor-insulator composites or
the permeability in viscous fluids flowing through porous media \cite{stauffer,sahimi}.
Unlike the universal (or quasi-universal) behavior of the critical exponents characterizing the
percolative transition, the value of the percolation threshold is a function of several variables
such as the shape of the percolating objects, their orientation and size dispersion, their
possible interactions and the microstructure in general \cite{torqua1}.

Of fundamental importance for several technological applications is the possibility of
exploiting such a multi-variable dependence to tailor the percolation threshold.
In particular, an issue of great interest concerns the problem of
lowering the percolation threshold, so to have long range connectivity of the percolating phase
with the minimum possible critical concentration. This is the case when, for objects dispersed in a continuous
medium, one wishes to exploit the properties of the percolating elements, but still preserving those of the
host medium. For example, low conducting filler amounts in a conductor-insulator composite
permit to obtain an adequate level of electrical conductivity with mechanical properties of
the composite being basically unaltered with respect to those of the pristine insulating phase.

There exist two main strategies to lower the percolation threshold through the manipulation of
the microstructure of heterogeneous composites. In one of such methods, one exploits the large
excluded volume $v_{\rm ex}$ of particle fillers with large aspect-ratios, such as rods
and/or disks dispersed in a three dimensional continuum medium \cite{balb1}, whose critical concentrations,
being proportional to $1/v_{\rm ex}$, can be made extremely small for sufficiently large aspect-ratios \cite{balb2}.
Such percolation threshold lowering has been studied in detail for several particle shapes and inter-particle
interactions, and is now well documented \cite{binen}.

\begin{figure}[t]
    \includegraphics[width=0.4\textwidth,clip=true]{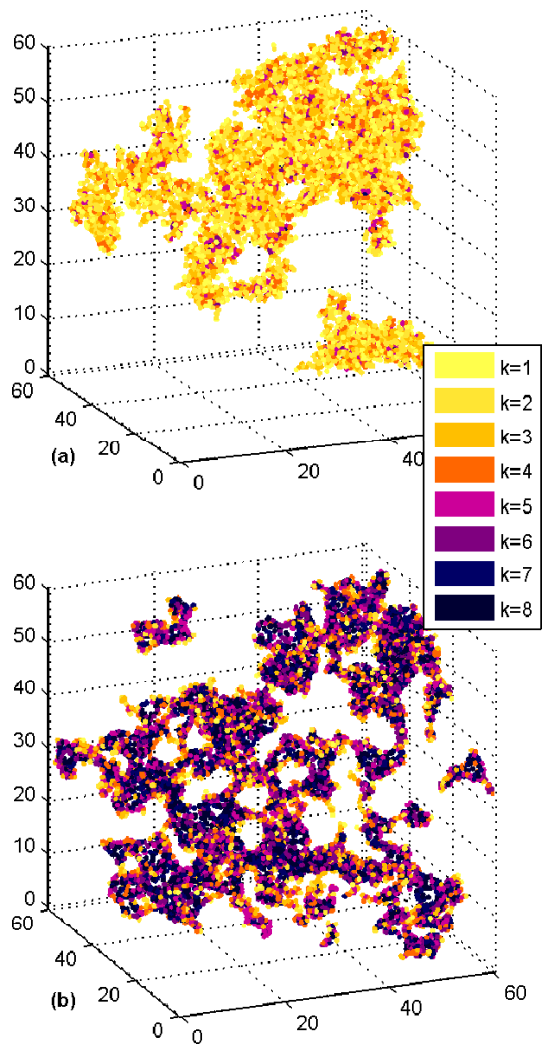}
        \caption{(Color) Percolating cluster of the conducting phase for (a) the homogeneous case $\phi_2=0$
        and for (b) the segregated regime with $\sigma_2/\sigma_1=12$ and $\phi_2=0.89$.
        The conducting particles are plotted together with their penetrable shells with $d=\sigma_1$.
        The color map defines the values of the connectivity number $k$ for each particle (see text).}
\label{fig1}
\end{figure}

A second strategy to lower the percolation threshold is obtained by forbidding the percolating objects
to occupy large (compared to the particle size) volumes inside the material, so to give rise to a segregated structure
like the one shown in Fig.~\ref{fig1}(b). In practice, this can be achieved when elements of two (mutually impenetrable)
species have different sizes and percolation is established by the smaller elements. Typical examples
of segregated systems are conductor-insulator composites where the size of the conducting
particles is much smaller than that of the insulating regions \cite{schueler,RuO2a,RuO2b}, which
display critical concentrations of a few percent or lower.

Despite that such low percolation thresholds are qualitatively understood by the reduced available volume for
arranging the conducting particles, very few studies exist on segregated percolation
in the continuum \cite{ekere}, while the vast majority of studies are limited to lattice representations
of the segregated structure \cite{RuO2a,malliaris}, which provide
only a partial understanding of the percolation properties of segregated systems.

In this Letter we consider a realistic continuum model of segregated percolation, primarily aimed at describing
the microstructure of segregated conductor-insulator composites, but general enough to represent
also other structurally similar systems such as particle-laden foams \cite{addad} or
filled asphaltene matrices \cite{wilbrink}.
We show that, by varying the degree of segregation of the system,
the percolation threshold is generally not a monotonous decreasing function of segregation, as suggested by earlier
studies \cite{RuO2a,ekere,malliaris}, but rather it displays a minimum before maximum segregation is reached.
Hence, the optimal percolation threshold does not necessarily coincide with the most
segregated structure, leading to a more complex phenomenology than previously thought.

We model a continuum segregated composite by considering one kind
of impenetrable spherical particles of diameter $\sigma_1$, which
may refer to the conducting objects in a conductor-insulator
composite, and a second kind of (insulating) spherical particles
with diameter $\sigma_2\geq \sigma_1$, which we allow to
penetrate each other. Furthermore, to generate segregation, we
assume that the two species of particles are mutually
impenetrable, and that the voids left over from the two kind of
particles are filled by the second (i.e., insulating) phase.
Finally, the connectivity criterion for the conducting phase is
defined by introducing a penetrable shell of thickness $d/2$
surrounding each conducting sphere, so that two given particles are
connected if their penetrable shells overlap.
This model represents a rather faithful description of real
segregated composites, such as the RuO$_2$-glass systems \cite{RuO2a,RuO2b}, where
thermal treatments on mixtures of RuO$_2$ and glassy grains
lead to composites made of conducting RuO$_2$ particles dispersed
in a continuum insulating glassy phase. Segregation is induced by
the larger size of the original glassy grains compared to that of
the conducting particles. Furthermore, in this and other similar classes
of composites, electrical transport is given by direct tunneling
or hopping processes, defining a characteristic length, represented
by $d$ in our model, below which two conducting particles are
electrically connected. Finally, it is worth to mention that the model
introduced here is relevant also for studying transport of macromolecules
in disordered porous media \cite{kim}, where $d$ represents in this case
the size of a test macromolecule and $\sigma_2/\sigma_1$ the pore size ratio
in a bi-dispersed porous medium.

In our numerical simulations, the system described above is
generated by first placing randomly the insulating spheres in a
cube of edge length $L$ with a given number density
$\rho_2=N_2/L^3$, where $N_2$ is the number of spheres. The
corresponding volume fraction for $L\rightarrow \infty$
is $\phi_2=1-\exp(-v_2\rho_2)$,
where $v_2=\pi\sigma_2^3/6$ is the volume of a single insulating
sphere. In a second step, $N_1$ conducting (and impenetrable)
particles of diameter $\sigma_1$ and number density
$\rho_1=N_1/L^3$ are added in the remaining void space and a
Metropolis algorithm is used to attain equilibrium \cite{torqua1}.
In the following, for the conducting phase, we use the reduced concentration
variable $\eta_1=\rho_1\pi/6(\sigma_1+d)^3$.
Examples of the resulting spatial distributions of the conducting
particles are shown in Fig.~\ref{fig1}(a) for the homogeneous case ($\phi_2=0$)
and in Fig.~\ref{fig1}(b) for a segregated one ($\sigma_2/\sigma_1=12$ with $\phi_2=0.89$)
for $\eta_1$ values close to their respective percolation thresholds (see below).

To obtain the critical density $\eta_1^c$ for infinitely large
system sizes we follow a standard finite-size scaling method. Namely,
for given values of $\sigma_2/\sigma_1$, $d$, and $\rho_2$,
and by using a modified Hoshen-Kopelman algorithm \cite{hoshen},
we calculate as a function of $\eta_1$ and $L$ the probability
$P(\eta_1,L)$ that a cluster of phase $1$ spans the system in a
given direction, with periodic boundary conditions in the other
two directions. The critical density $\eta_1^c(L)$ for finite $L$
is then extracted from the condition $P(\eta_1,L)=1/2$
\cite{ziff}, and $\eta_1^c$ follows from the scaling relation
$\eta_1^c(L)-\eta_1^c \propto L^{-1/\nu}$, where $\nu$ is the
correlation length exponent obtained from the width of the
transition. We considered $8$ different system sizes ranging from
$L=16$ with $N_{s}=1500$ realizations to $L=60$ ($N_s=100$) for $\sigma_2/\sigma_1=1$
and from $L=60$ ($N_s=200$) to $L=140$ ($N_s=100$) for $\sigma_2/\sigma_1=12$.
$20$ values of $\eta_1$ were typically used to
fit $P(\eta_1,L)$ with an appropriate function. In this way, for
most of the cases studied, the calculated $\nu$ values were well within $5\%$
of the universal value $\nu\simeq 0.88$ \cite{stauffer}.

\begin{figure}[t]
\protect
    \includegraphics[width=0.4\textwidth,clip=true]{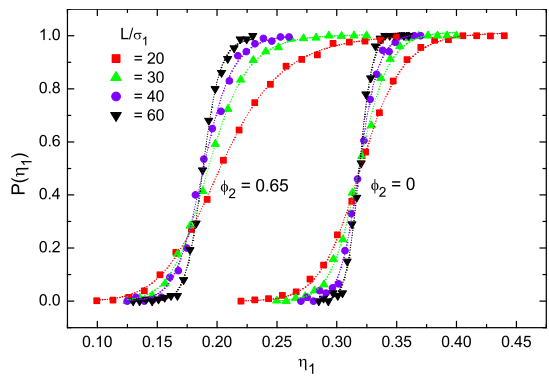}
        \caption{(Color online) Spanning probability as a function of
        $\eta_1$ for few values of the system linear size $L$ and for two
        different values of insulating phase volume fraction $\phi_2$.
        The penetrability length is $d=\sigma_1$ and $\sigma_2/\sigma_1=4$.}
        \label{fig2}
\end{figure}

Typical spanning probability results are reported in
Fig.~\ref{fig2}, where we plot $P(\eta_1,L)$ for
$\sigma_2/\sigma_1=4$, $d=\sigma_1$, and for two values of
$\phi_2$ with few different system sizes. As it is clear from the
figure, compared to the homogeneous case $\phi_2=0$, the spanning
probability transition for $\phi_2\neq 0$ gets shifted to lower
values of $\eta_1$, indicating that the percolation threshold is
reduced by segregation. This is confirmed by the scaling analysis
described above, which gives $\eta_1^c=0.3203\pm 0.0003$ for
$\phi_2=0$, which is in very good accord with Ref.\cite{heyes,Johner08},
and $\eta_1^c=0.1821\pm 0.0004$ for $\phi_2=0.65$.

\begin{figure}[t]
\protect
    \includegraphics[width=0.5\textwidth,clip=true]{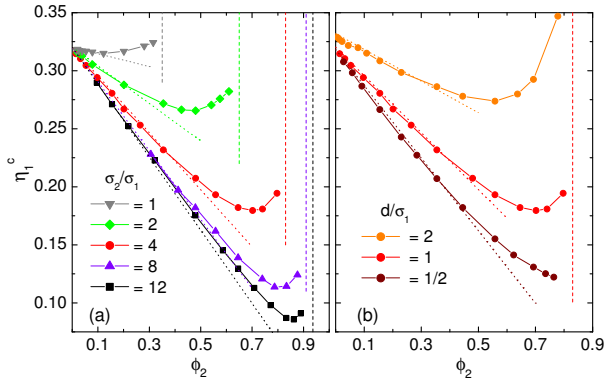}
        \caption{(Color online) Percolation threshold values $\eta_1^c$ as a
        function of the volume fraction $\phi_2$ of the insulating spheres for (a)
        $d=\sigma_1$ and several value of $\sigma_2/\sigma_1$ and (b) $\sigma_2/\sigma_1=4$
        and few values of $d$. The vertical dashed lines are lower bounds of the maximum
        segregation obtained from Eq.\eqref{void}, while the dotted lines are from Eq.\eqref{ansatz}.}
\label{fig3}
\end{figure}

Although the reduction of $\eta_1^c$ shown in Fig.~\ref{fig2} has to be expected on the basis
of reduced available volume arguments, we find that, actually, this is not always the case,
and that $\eta_1^c$ is generally a non-monotonous function of $\phi_2$ for
fixed $\sigma_2/\sigma_1$. This is shown in Fig.~\ref{fig3}(a) where $\eta_1^c$ is
plotted as a function of $\phi_2$ for $d=\sigma_1$ and for several values
of $\sigma_2/\sigma_1$, and in Fig.~\ref{fig3}(b) where $\sigma_2/\sigma_1=4$
and $d$ is varied. For all cases studied, on enhancing $\phi_2$ from the
homogeneous case $\phi_2=0$, the behavior of the percolation
threshold is characterized by an initial linear decrease of $\eta_1^c$, followed by
a minimum at a particular value of $\phi_2$ which depends
upon $\sigma_2/\sigma_1$ and $d$, and a final increase well before maximum segregation
is reached at $\phi_2^*$. Lower bounds of $\phi_2^*$ are plotted in Fig.~\ref{fig3}
by vertical dashed lines, which are obtained by requiring that the available
volume fraction for arranging the conducting particles
\begin{equation}
\label{void}
\phi_{\rm avail.}=(1-\phi_2)^{(1+\sigma_1/\sigma_2)^3}
\end{equation}
coincides with the percolating
volume fraction of voids, $\phi_{\rm void}^c$, which for the three dimensional
penetrable sphere model used here is $\phi_{\rm void}^c\simeq 0.03$ \cite{ker}.

As shown in Fig.~\ref{fig3}(a), the slope of the initial decrease of $\eta_1^c$
is steeper for $\sigma_2/\sigma_1$ larger, and the position of the minimum
gets shifted to higher values of $\phi_2$. A similar effect is found by decreasing the
penetrable shell thickness $d$ for fixed $\sigma_2/\sigma_1$, Fig.~\ref{fig3}(b),
leading to infer that for $d/\sigma_2\rightarrow 0$ the minimum disappears and
$\eta_1^c$ decreases monotonously all the way up to $\phi_2^*$.
These features, and in particular the appearance of a minimum (i.e. optimal) value
of the percolation threshold for finite penetrable shells, represent our main finding and
provide a previously unnoticed
scenario for segregated percolation.

Let us discuss now the physical origin of the non-monotonous behavior
of the percolation threshold.
The initial decrease of $\eta_1^c$ can be fairly well reproduced by assuming that,
for low values of $\phi_2$, the volume fraction $\phi_1^c$ of the composite conducting particles
(hard-core plus penetrable shell) is reduced by the volume occupied by
insulating spheres. However, since the penetrable shells of the conducting particles may
actually overlap the insulating spheres, these latter may be treated as having
effectively a smaller volume, $v_{\rm eff}\leq v_2$, leading to
$\phi_1^c(\phi_2)\simeq\phi_1^c(0)(1-\phi_2 v_{\rm eff}/v_2)$. Taking into account that
insulating particles with $\sigma_2\lesssim a$, where $a$ is the mean distance between
the closest surfaces of nearest neighbor conducting particles, should be ineffective in reducing
$\phi_1^c$, we approximate $v_{\rm eff}$ by a sphere of diameter $\sigma_2-a$. Finally,
by expanding $\phi_1^c(\phi_2)$ in powers of $\eta_1^c(\phi_2)-\eta_1^c(0)$,
at the lowest order in $\phi_2$ we find
\begin{equation}
\label{ansatz}
\eta_1^c(\phi_2)\simeq\eta_1^c(0)-\frac{\phi_1^c(0)}{\phi_1^c(0)'}\left(\frac{\sigma_2-a}{\sigma_2}\right)^3\phi_2,
\end{equation}
where $\phi_1^c(0)'=\lim_{\phi_2\rightarrow 0} d\phi_1^c(\phi_2)/d\eta_1$.
As it is seen in Fig.~\ref{fig3}, where Eq.\eqref{ansatz} (dotted lines) is plotted by
using $a=(\sigma_1+d)/2\eta_1^c(0)^{1/3}-\sigma_1$ \cite{Johner08} and $\phi_1(0)$
as given in Ref.\cite{torqua1}, the low $\phi_2$ behavior of $\eta_1^c$ is rather well
reproduced for all cases considered.

\begin{figure}[t]
\protect
    \includegraphics[width=0.4\textwidth,clip=true]{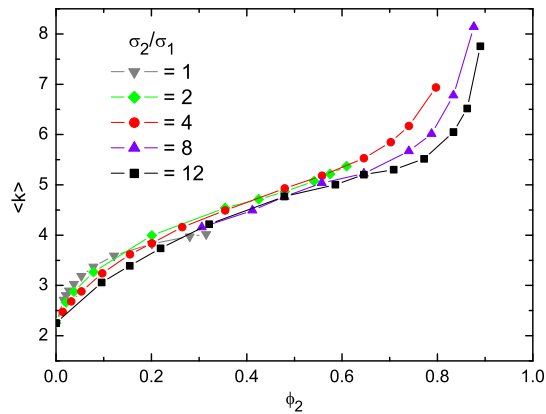}
        \caption{(Color online) Mean connectivity number $\langle k\rangle$ as a function of $\phi_2$ for the
        same cases of Fig.~\ref{fig3}(a).}
\label{fig4}
\end{figure}

By construction, the above argument neglects possible effects of $\phi_2\neq 0$ on the
connectivity number $k$, i.e., the number of conducting particles directly connected to a given one.
Actually, as it is shown in Fig.~\ref{fig1} where the color map defines $k$ for each particle in the
percolating cluster, the rather narrow $k$-distribution for the homogeneous case, which
is peaked around the mean value $\langle k\rangle\simeq 2.25$ \cite{heyes,binen},
changes drastically in the highly segregated regime of Fig.~\ref{fig1}(b). Here, clusters of highly
connected particles ($k$ large) are bound together by ``chains'' of particles having
low $k$ values. Such distribution of $k$ values is due to the fact that, in the vicinity of $\phi_2^*$, the structure
of the void space available for arranging the centers of the conducting particles is characterized by many narrow
(quasi-one dimensional) necks connecting more extended void regions \cite{halperin}.
Percolation is possible only if such necks are populated by connected conducting particles,
and since for $\phi_2 \rightarrow \phi_2^*$ the necks become narrower, and so have less probability
of being populated, more particles are needed
to ensure connectivity, thereby ``overcrowding'' the many void regions between the necks. The net
effect of such mechanism, not captured by Eq.\eqref{ansatz}, is the enhancement of $\eta_1^c$ as
$\phi_2 \rightarrow \phi_2^*$. This is demonstrated in Fig.~\ref{fig4} where $\langle k\rangle$,
plotted for the same cases of Fig.~\ref{fig3}(a), displays a sudden enhancement (more marked for
$\sigma_2/\sigma_1$ larger) at values of $\phi_2$
corresponding to the points of upturn of $\eta_1^c$ of Fig.~\ref{fig3}(a).
The competition between the effect of reduced available volume, which lowers
$\eta_1^c$ [Eq.\eqref{ansatz}], and the enhanced connectivity at high segregation, which increases
$\eta_1^c$, is therefore at the origin of the minimum percolation threshold observed by us.

Before concluding, let us discuss the possibility of observing the features presented here in
real segregated materials. In conductor-insulator composites where transport is driven by tunneling,
$d$ represents the maximum tunneling distance between the conducting particles, so that $d$
would be of the order of few nanometers. The same is expected for particles linked
by organic molecules, as discussed in Ref.\cite{mueller}. For such values of $d$,
the results of Fig.~\ref{fig3} would therefore apply
to nano-composites with $\sigma_1\approx d$ and $\sigma_2$ not exceeding a few tens of nanometers.
Much larger values of $d$ are however possible in some RuO$_2$-glass composites, where a reactive
layer of thickness $0.2$-$0.4$ $\mu$m (or even more) surrounding the RuO$_2$ particles presents modified chemical
and structural properties \cite{adachi}, most probably favoring hopping processes \cite{mene}.
In this case, the parameters used in our work
would easily account for composites with $\sigma_1$ in the range $50$-$500$ nm and $\sigma_2$ of
few microns.

This work was supported by the Swiss National Science Foundation (Grant No. 200020-116638).
We thank G. Ambrosetti and I. Balberg for valuable discussions.


\begin{thebibliography}{99}

\bibitem{stauffer}
D. Stauffer and A. Aharony, \textit{Introduction to Percolation Theory} (Taylor \& Francis, London, 1992).

\bibitem{sahimi}
M. Sahimi, \textit{Heterogeneous Materials I} (Springer, New York, 203).

\bibitem{torqua1}
S. Torquato, \textit{Random Heterogeneous Materials: Microstructure and Macroscopic Properties}
(Springer, New York, 2002).

\bibitem{balb1}
I. Balberg, C. H. Anderson, S. Alexander, and N. Wagner,
Phys. Rev. B {\bf 30}, 3933 (1984).

\bibitem{balb2}
I. Balberg, Phys. Rev. B {\bf 33}, 3618(R) (1986).

\bibitem{binen}
E. Charlaix, J. Phys. A: Math. Gen. {\bf 19}, L533 (1986);
I. Balberg and N. Binenbaum, Phys. Rev. A {\bf 35}, 5174 (1987);
E. J. Garboczi, K. A. Snyder, J. F. Douglas, and M. F. Thorpe,
Phys. Rev. E {\bf 52}, 819 (1995);
T. Schilling, S. Jungblut, and M. A. Miller, Phys. Rev. Lett. {\bf 98}, 108303 (2007).

\bibitem{schueler}
R. Schueler, J. Petermann, K. Schute, and H.-P. Wentzel,
J. Appl. Polym. Sci. {\bf 63}, 1741 (1997);
W. J. Kim, M. Taya, K. Yamada, and N. Kamiya, J. Appl. Phys. {\bf 83}, 2593 (1998);
C. Chiteme and D. S. McLachlan, Phys. Rev. B {\bf 67}, 024206 (2003).

\bibitem{RuO2a}
A. Kubovy, J. Phys. D: Appl. Phys. {\bf 19}, 2171 (1986);
A. Kusy, Physica B {\bf 240}, 226 (1997).

\bibitem{RuO2b}
P. F. Carcia, A. Ferretti, and A. Suna, J. Appl. Phys. {\bf 53}, 5282 (1982);
S. Vionnet-Menot, C. Grimaldi, T. Maeder, S. Str\"assler, and P. Ryser,
Phys. Rev. B {\bf 71}, 064201 (2005).

\bibitem{ekere}
A. S. Ioselevich and A. A. Kornyshev, Phys. Rev. E {\bf 65}, 021301 (2002);
D. He and N. N. Ekere, J. Phys. D: Appl. Phys. {\bf 37}, 1848 (2004).

\bibitem{malliaris}
A. Malliaris and D. T. Turner, J. Appl. Phys. {\bf 42}, 614 (1971);
R. P. Kusy, J. Appl. Phys. {\bf 48}, 5301 (1977);
I. J. Youngs, J. Phys. D: Appl. Phys. {\bf 36}, 738 (2003).

\bibitem{addad}
S. Cohen-Addad, M. Krzan, R. H\"ohler, and B. Herzhaft,
Phys. Rev. Lett. {\bf 99}, 168001 (2007)

\bibitem{wilbrink}
M. W. L. Wilbrink, M. A. J. Michels, W. P. Vellinga, and H. E. H. Meijer
Phys. Rev. E {\bf 71}, 031402 (2005).

\bibitem{kim}
I. C. Kim and S. Torquato, J. Chem. Phys. {\bf 96}, 1498 (1992).

\bibitem{hoshen}
J. Hoshen and R. Kopelman, Phys. Rev. B \textbf{14}, 3438 (1976).

\bibitem{ziff}
R. M. Ziff, Phys. Rev. Lett. {\bf 69}, 2670 (1992);
M. D. Rintoul and S. Torquato, J. Phys. A \textbf{30}, L585
(1997).

\bibitem{heyes}
D. M. Heyes, M. Cass, and A. C. Branca, Mol. Phys. \textbf{104},
3137 (2006).

\bibitem{Johner08}
N. Johner, C. Grimaldi, I. Balberg, and P. Ryser, Phys. Rev. B \textbf{77}, 174204 (2008).

\bibitem{ker}
J. Kertesz, J. Phys. Lett.-Paris \textbf{42}, L393 (1981);
W. T. Elam, A. R. Kerstein, and J. J. Rehr, Phys. Rev. Lett. \textbf{52}, 1516 (1984).

\bibitem{halperin}
S. Feng, B. I. Halperin, and P. N. Sen, Phys. Rev. B {\bf 35}, 197 (1987).

\bibitem{mueller}
K.-H. M\"uller, J. Herrmann, B. Raguse, G. Baxter, and T. Reda,
Phys. Rev. B {\bf 66}, 075417 (2002).

\bibitem{adachi}
K. Adachi, S. Iida, and K. Hayashy, J. Mater. Res. {\bf 9}, 1866 (1994).

\bibitem{mene}
C. Meneghini, S. Mobilio, F. Pivetti, I. Selmi, M. Prudenziati, and B. Morten,
J. Appl. Phys. {\bf 86}, 3590 (1999).

\end{thebibliography}
\end{document}